\documentclass[aps,prb]{revtex4}
\usepackage{bm}
\usepackage{amsmath}
\usepackage{color}
\usepackage{graphicx}
\usepackage{amssymb}
\DeclareMathAlphabet{\mathbbmsl}{U}{bbm}{m}{sl}
\makeatletter
\newsavebox{\@brx}
\newcommand{\llangle}[1][]{\savebox{\@brx}{\(\m@th{#1\langle}\)}%
	\mathopen{\copy\@brx\kern-0.5\wd\@brx\usebox{\@brx}}}
\newcommand{\rrangle}[1][]{\savebox{\@brx}{\(\m@th{#1\rangle}\)}%
	\mathclose{\copy\@brx\kern-0.5\wd\@brx\usebox{\@brx}}}
\makeatother

\begin{document}
\title{Quantum Hall Effect and Chern Phases in the 1/5-Depleted Square Lattice}
%\title{Quantum Hall effect on the 1/5-depleted square lattice}% Force line breaks with \\
\author{Sara Aghtouman$^1$}
\author{Godfrey Gumbs$^{2,3,4}$ }
\author{Mir Vahid Hosseini$^1$}
 \email[Corresponding author: ]{mv.hosseini@znu.ac.ir}
\affiliation{$^1$Department of Physics, Faculty of Science, University of Zanjan, Zanjan 45371-38791, Iran}
\address{$^2$Department of Physics, Hunter College, City University of New York, 695 Park Avenue, New York, NY 10065 USA}
\address{$^{3}$The Graduate School and University Center, The City University of New York, New York, NY 10016, USA}
\address{$^4$Donostia International Physics Center (DIPC), P de Manuel Lardizabal, 4, 20018 San Sebastian, Basque Country, Spain}

\date{\today}

\begin{abstract} 
We investigate the fractal energy spectrum and quantum Hall response of a two-dimensional 1/5-depleted square lattice subjected to a perpendicular magnetic field. Using a tight-binding model that includes both nearest-neighbor ($t_
1$) and next-nearest-neighbor ($t_2$) hopping, we compute the Hofstadter butterfly and extract quantized Hall conductivities via Chern number calculations. In the absence of diagonal hopping ($t_2=0$), the spectrum exhibits exact particle–hole and flux-inversion symmetries, and the total Chern number across all bands vanishes. When $t_2$ is introduced, these symmetries are broken, the butterfly becomes deformed, new gaps open, and—remarkably—a nonzero total Chern sum can emerge, signaling unconventional topological phases. By systematically varying $t_1$ and $t_2$, we identify regimes with large individual Chern indices and parameter windows where gap stability and Hall plateaus are optimized. Our results demonstrate that lattice depletion combined with diagonal hopping provides a tunable route to engineer robust Chern insulators in both artificial and oxide-based square-lattice systems.
\end{abstract}

%\pacs{ }
\maketitle

%%%%%%%%%%%%%%%%%%%%%%%%%%%%%%%%%%%%%%%%%%%%%%%%%%%%%%%%%%%%%%%%%%%%%%%%%%%
\section{Introduction} \label{s1}

Topological phases of matter have emerged as a central theme in condensed matter physics \cite{TI1,TI2}, characterized by global invariants that remain robust against local perturbations \cite{{TISymetry}}. The quantum Hall effect (QHE), first observed by von Klitzing \cite{QHE}, exemplifies such a phase, where the Hall conductance is quantized in integer multiples of $e^2/h$, corresponding to topologically invariant Chern numbers \cite{thouless}. This discovery laid the foundation for the classification of electronic phases based on topological properties rather than symmetry breaking \cite{Kohmo}. Subsequent theoretical advancements, such as Haldane's model of a QHE without Landau levels, introduced the concept of Chern insulators, further enriching the landscape of topological phases \cite{Haldane}. Moreover, recent studies have demonstrated that materials with higher Chern numbers can host quantum anomalous Hall effects \cite{QAHE}, offering avenues for dissipationless edge transport in systems without external magnetic fields. These developments underscore the rich interplay between topology and geometry from condensed matter systems \cite{Hallrev} to optical lattices \cite{4DHall}.

A seminal discovery in this context was made by Hofstadter, who found that in a square lattice subjected to a uniform magnetic field, Bloch electrons exhibit a fractal energy spectrum known as the Hofstadter butterfly \cite{hofstadter}. This rich structure emerges from the interplay between two incommensurate length scales: the magnetic length and the lattice periodicity, leading to miniband formation and quantized gaps in the spectrum \cite{harper,wannier}. Hofstadter’s butterfly could recently be experimentally observed \cite{HofExp1,HofExp2,HofExp3,HofExp4,HofExp5,HofExp6,HofExp7,HofExp8,HofExp9,HofExp10,HofExp11,HofExp12,HofExp13,HofExp14}. Subsequent theoretical studies have expanded Hofstadter magneto-bands to a wide range of geometries including triangular \cite{artificialgraphene}, honeycomb \cite{lattice1,lattice2}, kagome \cite{lattice3,lattice4,floquettriangular}, and dice lattices \cite{lattice5} with nonperturbative description \cite{nonperturbative}. Similar fractal characteristics have been investigated in systems such as monolayer graphene \cite{grapheneBN} under circularly polarized light \cite{floquetGraphene} and periodically modulated transverse magnetic fields \cite{monolayergraphene}, twisted bilayer graphene \cite{twist0,twist1,twist2,twist3,twist4,twist5,twist6,twist7}, quasicrystals \cite{quasi1}, ladder networks \cite{ladder}, and Sierpiński carpets and gaskets \cite{sier1,sier2,sier3}. Furthermore, the Hofstadter model has been extended to include vacancy defects within a lattice \cite{defect}. These findings cemented the Hofstadter model as a cornerstone for understanding quantum Hall physics in lattice systems, revealing how topologically nontrivial minibands can emerge under magnetic modulation \cite{MagModu}.

Beyond traditional settings with real magnetic fields, Hofstadter spectra can be enriched in systems via periodic modulation of lattice parameters \cite{SSH2D1,SSH2D2}, such as hopping amplitudes, sublattice potentials, or on-site energies. Notably, periodic modulation of lattice parameters can also generate synthetic gauge fields that mimic the effects of magnetic flux without requiring an external field \cite{ssh1D}. These approaches have enabled the observation of Hofstadter butterfly structures in cold-atom optical lattices \cite{optical1,optical2}, with engineered artificial magnetic fields \cite{hafezi2013}, and electronic systems with moiré superlattices \cite{moire1,moire2}, where spatial variation in stacking creates an emergent flux landscape. Recent studies on topological phases in quasiperiodic modulations of hoppings in two-leg ladder quasicrystals \cite{1Dquasi1} and the interplay of internal degrees of freedom in non-Abelian gauge field models \cite{nonabelian} have been shown to produce fractal energy spectra reminiscent of the Hofstadter butterfly. These advances demonstrate that fractal spectral structures and topologically nontrivial bands are not exclusive to real magnetic fields, but can also be engineered through geometric design and synthetic fields—greatly expanding the platform for realizing topological quantum phenomena.

Despite extensive research on regular square lattices, depleted square lattices—such as the 1/5-depleted configuration—remain relatively unexplored.
This geometry naturally gives rise to a pattern of large and small square plaquettes. This configuration is particularly compelling because the depletion creates a unique interplay between nearest-neighbour (NN) and next-nearest-neighbour (NNN) hoppings, which in turn may give rise to rich topological phenomena. Notably, the real material CaV$_4$O$_9$, which exhibits the 1/5-depleted square lattice \cite{depleted1,depleted2,depleted3,depleted4}, has attracted considerable attention due to its novel physical properties \cite{square1}. Quantum phase transitions that elucidate magnetic properties have been induced \cite{quantumphase1,quantumphase2}, and critical phenomena associated with these transitions have also been investigated \cite{phasetransition1,phasetransition2,phasetransition3}.

In this work, we investigate the energy spectrum and quantum Hall effects of the 1/5-depleted square lattice subjected to a uniform magnetic flux applied selectively through either the large or small square plaquettes, reminiscent of a dot–antidot lattice system \cite{DotAntodot}. Using a tight-binding model that includes both NN and NNN hopping terms, we analyzed the resulting Hofstadter butterfly spectrum and the associated quantized Hall conductivities. Our results demonstrate that varying the hopping parameters not only reshapes the band structure but also induces topological phase transitions, characterized by changes in the distribution and magnitude of Chern numbers. Notably, we identified regimes where NNN hopping breaks spectral symmetries and leads to nonzero net Chern sums—highlighting a pathway to unconventional topological states in lattice systems.

\section{Model and Theory}\label{s2}
%%%%%%%%%%%%%%%%%%%%%%%%%%%%%%%%%%%%%%%%%%%%%%%%%%%%%%%%%%%%%%%%%%%%%%%%%%%

\begin{figure}[t!] 
\centering 
\includegraphics[width=7.5cm]{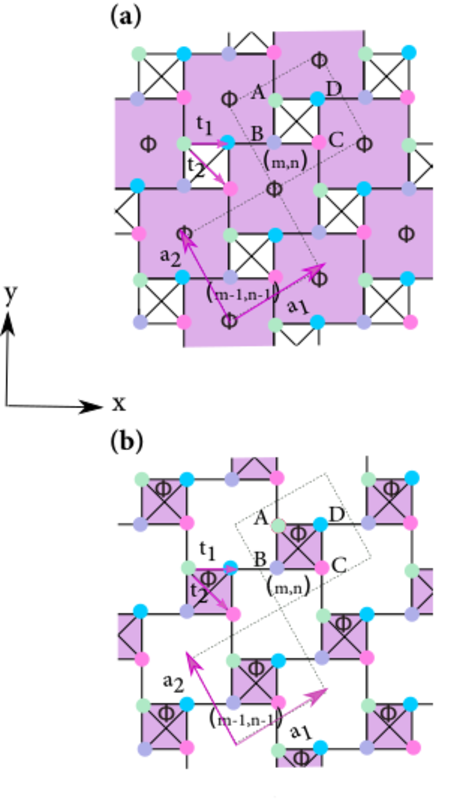} 
\caption{(Color online) Schematic diagram of a 1/5-depleted square lattice model. The black dotted lines indicate the unit cell, which contains four sites labeled A, B, C, and D. The nearest-neighbor (NN) hopping parameter along the sides of the small square is denoted by $t_{1}$, and the next-nearest-neighbor (NNN) hopping parameter along the diagonals is denoted by $t_{2}$. Both the large square in (a) and the small square in (b) are threaded by a uniform external magnetic flux $\Phi$.} 
\label{fig1} 
\end{figure}

We consider a two-dimensional 1/5-depleted square lattice subjected to a perpendicular magnetic field that includes two patterns of magnetic field penetration. Figures~\ref{fig1}(a) and \ref{fig1}(b) schematically illustrate a two-dimensional 1/5-depleted square lattice with an external magnetic flux, $\Phi$, penetrating the large and small squares, respectively. We adopt the Landau gauge $A=(0,Bx,0)$ which produces a uniform magnetic field on the xy-plane. Under the influence of the vector potential, the hopping parameter will be modified according to the Peierls substitution, i.e., $t\rightarrow te^{ i\frac{e}{\hbar}\int_{i}^{j}A\cdot dl}$. Each unit cell contains four sublattices labeled A, B, C, and D with the primitive lattice vectors $a_{1}=(2,1)$ and $a_{2}=(-1,2)$, so the position of a unit cell is given by $R(m,n)=ma_{1}+na_{2}$.

This system can be modeled by a tight-binding Hamiltonian, that includes both NN and NNN hoppings, as \cite{quantumphase2,hamil} 
\begin{align}\label{eq1}
    H&=\sum_{m,n}[t_{1}(e^{i\alpha_{m}}c^{\dagger}_{m,n,B}c_{m,n,A}+e^{-i\alpha_{m}}c^{\dagger}_{m,n+1,C}c_{m,n,A}\\\nonumber&+c^{\dagger}_{m,n,D}c_{m,n,A}+c^{\dagger}_{m+1,n,B}c_{m,n,D}+e^{i\alpha_{m}}c^{\dagger}_{m,n,C}c_{m,n,D}\\\nonumber&+c^{\dagger}_{m,n,B}c_{m,n,C}+c^{\dagger}_{m-1,n,D}c_{m,n,B}+e^{i\alpha_{m}}c^{\dagger}_{m,n-1,A}c_{m,n,C})\\\nonumber&+t_{2}(e^{i\beta_{m}}c^{\dagger}_{m,n,C}c_{m,n,A}+e^{i\beta_{m}}c^{\dagger}_{m,n,B}c_{m,n,D})+H.c]  
\end{align}
where the summation over $(m,n)$ runs over the unit cell indices and $c^{\dagger}_{m,n,i}$ ($c_{m,n,i}$) is the creation (annihilation) operator for an electron at the $i$-th site in the $(m,n)^\text{th}$ unit cell. Here, $t_{1}$ denotes the NN hopping parameter along the boundary of a small square plaquette, and $t_{2}$ denotes the NNN hopping parameter along the diagonal directions. The phases $\alpha_m$ and $\beta_m$, originating from the Peierls substitution, are  
\begin{equation}
   \alpha_{m}=-\pi\frac{4f}{S}m, \quad\beta_{m}=-\pi\frac{4f}{S}(m+\frac{1}{4}).
\end{equation}
For a rational flux, we have taken the flux per unit cell to be $f=\Phi/\Phi_{0}=p/q$, where $\Phi=SB$ is the magnetic flux through the unit cell with the area $S$ and $\Phi_{0}=\frac{h}{e}$ is the quantum magnetic flux with $h$ being the Planck constant and $e$ the electron charge. $p$ and $q$ are co-prime integers and $S$ is the area of the unit cell pierced by magnetic flux, e.g., $S=4$ for the large squares in Fig. \ref{fig1}(a) and $S=1$ for the small squares in Fig. \ref{fig1}(b). 

As described above, the magnetic flux \(\Phi\) is applied through a unit cell of area \(S\), so the effective magnetic field is given by \(B = f \Phi_0 / S\). This relation highlights an important distinction between the two flux patterns illustrated in Fig. \ref{fig1}(a) and \ref{fig1}(b): when flux is threaded through the large plaquettes (\(S = 4\)), the resulting magnetic field is four times weaker than when the same flux fraction \(f\) is applied to the small plaquettes (\(S = 1\)). Since both the cyclotron radius \(r_c \propto 1/B\) and the magnetic length \(\ell_B = \sqrt{\hbar / eB} \propto \sqrt{S/f}\) are inversely related to the magnetic field strength, it follows that magnetic orbits are larger and more delocalized in the large-plaquette configuration and tighter and more localized in the small-plaquette case. Physically, this means that in the large-plaquette geometry, the magnetic modulation varies more slowly over the lattice scale, resulting in flatter minibands and smoother Hofstadter features. In contrast, threading flux through smaller plaquettes produces a sharper magnetic potential landscape, which enhances Landau-level splitting and opens more prominent energy gaps. This distinction plays a central role in shaping the Hofstadter spectrum and the associated distribution of Chern numbers in the two geometries.

\begin{figure*}[th!]
    \centering
    \includegraphics[width=14cm]{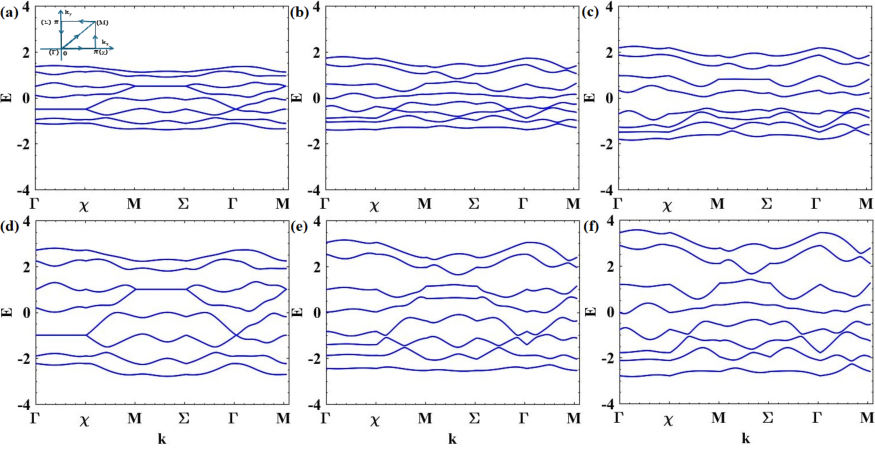}
    \caption{Band structure for periodic boundary conditions for large squares threaded by magnetic flux at \(f=1\) for various combinations of \(t_{1}\) and \(t_{2}\): (a) \(t_{1}=0.5\), \(t_{2}=0\); (b) \(t_{1}=0.5\), \(t_{2}=0.5\); (c) \(t_{1}=0.5\), \(t_{2}=1\); (d) \(t_{1}=1\), \(t_{2}=0\); (e) \(t_{1}=1\), \(t_{2}=0.5\); (f) \(t_{1}=1\), \(t_{2}=1\). Inset: First Brillouin zone of the 1/5-depleted square
lattice.}
    \label{band}
\end{figure*}

Assuming the lattice translation symmetry along the $y$ direction so that the crystal momentum $k_{y}$ is a good quantum number and applying Bloch's boundary conditions in the x direction, the Hamiltonian can be written as 
\begin{equation}
H=\sum_{\textbf{k}}C^{\dagger}_{\textbf{k}}H_{\textbf{k}}C_{\textbf{k}},
     \label{eq2}
\end{equation}
where the summation is over the magnetic Brillouin zone defined by 
\begin{equation}
    -\frac{\pi}{qa_{1x}}\leq k_{x}\leq\frac{\pi}{qa_{1x}}, \quad -\frac{\pi}{a_{2y}}\leq k_{y}\leq\frac{\pi}{a_{2y}},
\end{equation}
and 
\begin{equation} 
C^{\dagger}_{\mathbf{k}} = \Big(c^{(1)\dagger}_{A,\mathbf{k}}, c^{(1)\dagger}_{B,\mathbf{k}}, c^{(1)\dagger}_{C,\mathbf{k}}, c^{(1)\dagger}_{D,\mathbf{k}}, \dots, c^{(q)\dagger}_{A,\mathbf{k}}, c^{(q)\dagger}_{B,\mathbf{k}}, c^{(q)\dagger}_{C,\mathbf{k}}, c^{(q)\dagger}_{D,\mathbf{k}}\Big).
\label{eq3} 
\end{equation}
The matrix representation of Hamiltonian in the momentum space, $H_{\textbf{k}}$, being a $4q\times4q$ matrix, can be obtained as
\begin{equation}
\begin{aligned}[b]
H_{\textbf{k}}=&\begin{pmatrix}E^{(1)}_{\textbf{k}}&F^{\dagger}&0&\dots&\dots&\dots&G_{\textbf{k}}\\F&E^{(2)}_{\textbf{k}}&F^{\dagger}&0&\dots&\dots&0\\0&F&E^{(3)}_{\textbf{k}}&F^{\dagger}&\dots&\dots&0\\\vdots&\ddots&\ddots&\ddots&\ddots&\ddots&\vdots\\0&\dots&0&F&E^{(q-2)}_{\textbf{k}}&F^{\dagger}&0\\0&\dots&\dots&0&F&E^{(q-1)}_{\textbf{k}}&F^{\dagger}\\G^{\dagger}_{\textbf{k}}&\dots&\dots&\dots&0&F&E^{(q)}_{\textbf{k}}\end{pmatrix}.
\label{eq4}
\end{aligned}
\end{equation}
Here,
\begin{equation}
\begin{aligned}[b]
E^{(m)}_{\textbf{k}}=&\begin{pmatrix}0&\eta_{m}&\zeta_{m}&t_{1}\\\eta^{*}_{m}&0&t_{1}&\kappa_{m}\\\zeta^{*}_{m}&t_{1}&0&\eta^{*}_{m}\\t_{1}&\kappa^{*}_{m}&\eta_{m}&0\end{pmatrix},
F=&\begin{pmatrix}0&0&0&0\\0&0&0&t_{1}\\0&0&0&0\\0&0&0&0\end{pmatrix},\\G_{\textbf{k}}=&Fe^{-iqk_{x}},
 \label{eq5}
\end{aligned}
\end{equation}
where 
\begin{equation}
\begin{aligned}[b]
\eta_{m}=&t_{1}e^{i\alpha_{m}},\quad \kappa_{m}=t_{2}e^{-i\beta_{m}},
\\\zeta_{m}=&t_{1}e^{i(ky-\alpha_{m})}+t_{2}e^{i\beta_{m}},
 \label{eq6}
\end{aligned}
\end{equation}
with $m=1,...,q$. 

The eigenvalue equation, 
\begin{equation}
H_{\textbf{k}}u^{n}_{\textbf{k}}=\epsilon^{n}_{\textbf{k}}u^{n}_{\textbf{k}},    
\end{equation}
can be solved, yielding the single-particle energies $\epsilon^{n}_{\textbf{k}}$ and eigen functions $u^{n}_{\textbf{k}}$ for the $n^{th}$ band. The resulting spectrum for varying flux $f=p/q$ $(p=1,...,q)$ is known as Hofstadter’s butterfly \cite{hofstadter}. 

Moreover, we computed the Hall conductivity to elucidate the system's topological properties. The Hall conductivity is given by 
\begin{equation} 
\sigma_{H} = -\frac{e^{2}}{h} C_{n}, 
\label{eq7} 
\end{equation} 
where $C_{n}$ denotes the Chern number associated with the $n^{th}$ band. The Chern number $C_{n}$ is as,
\begin{equation}
   C_{n}=\frac{1}{2\pi i}\int_{FBZ} d^{2}\textbf{k} \Omega_{n}(k_{x},k_{y}),
    \label{eq8}
\end{equation}
which can be computed numerically by discretizing the Brillouin zone \cite{Fuku}. Here, $\Omega_{n}=(\langle\partial_{k_{x}}u^{n}_{\textbf{k}}|\partial_{k_{y}}u^{n}_{\textbf{k}}\rangle-\langle\partial_{k_{y}}u^{n}_{\textbf{k}}|\partial_{k_{x}}u^{n}_{\textbf{k}}\rangle )$ is the Berry curvature. Note that for a rational flux $f=p/q$, each gap labeled with an integer $r$ is characterized by two topological invariants, $(C_{r},s_{r})$, satisfying the Diophantine equation \cite{gold,kohm,macdon}
\begin{equation}
   r=pC_{r}+qs_{r}.
    \label{eq9}
\end{equation}

%%%%%%%%%%%%%%%%%%%%%%%%%%%%%%%%%%%%%%%%%%%%%%%%%%%%%%%%%%%%%%%%%%%%%%%%%%%
\section {Results and discussion} \label{s3}
%%%%%%%%%%%%%%%%%%%%%%%%%%%%%%%%%%%%%%%%%%%%%%%%%%%%%%%%%%%%%%%%%%%%%%%%%%%  

\subsection{Band structure at the maximum flux}

\begin{figure*}[th!]
    \centering
    \includegraphics[width=14cm]{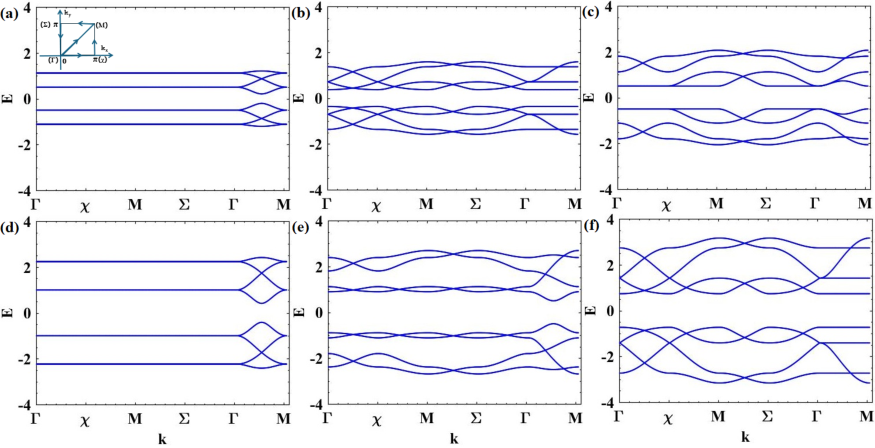}
    \caption{Band structure for periodic boundary conditions for small squares threaded by magnetic flux at \(f=1\) for various combinations of \(t_{1}\) and \(t_{2}\): (a) \(t_{1}=0.5\), \(t_{2}=0\); (b) \(t_{1}=0.5\), \(t_{2}=0.5\); (c) \(t_{1}=0.5\), \(t_{2}=1\); (d) \(t_{1}=1\), \(t_{2}=0\); (e) \(t_{1}=1\), \(t_{2}=0.5\); (f) \(t_{1}=1\), \(t_{2}=1\). Inset: First Brillouin zone of the 1/5-depleted square
lattice.}
    \label{baand}
\end{figure*}

In the following, we analyze the behavior of the band structure with different values of the NN hopping parameter \(t_{1}\) and the NNN hopping parameter \(t_{2}\) at $f=1$ under periodic boundary conditions. The band structures of the lattice patterns shown in Fig.~\ref{fig1}(a) (large squares) and Fig.~\ref{fig1}(b) (small squares) are presented in Figs. \ref{band} and \ref{baand}, respectively.

Figure \ref{band} illustrates how the band structure of the 1/5-depleted square lattice reorganizes under periodic boundary conditions when only the large plaquettes ($S=4$) are threaded by magnetic flux at $f=1$, for six representative ($t_1$, $t_2$) combinations. In panel (a), only weak NN hopping is present. The resulting spectrum is compressed in energy, with most band gaps at high-symmetry points remaining closed. Only tiny gap windows appear near the spectrum's upper and lower edges. Physically, this means that moderate edge‐path hopping alone is insufficient to open robust gaps in the large‐square geometry at full flux—electronic states remain largely overlapping and delocalized. In panel (b) the introduction of moderate NNN hopping dramatically alters the picture. Even at half‐strength $t_2$, multiple mid‐spectrum gaps appear that were entirely closed in panel (a). These new gaps arise because the additional diagonal paths provide extra Peierls phase loops, breaking the perfect bipartite interference of the NN‐only lattice and separating bands that previously overlapped. In panel (c), the NNN hopping pathways now dominate the interference network. The spectrum “fans out,” and numerous wide gaps emerge across the central energy region. So, the bands become well‐isolated from one another. Electrons acquire significant phase differences along diagonal loops, which enhance localization and yield clearly separated bands—hallmarks of a topologically active regime.

In panel (d), doubling the NN hopping broadens the overall bandwidth, most of the symmetry-protected gaps remain closed.  However, the lack of diagonal hopping means no new topological gaps emerge, highlighting the limited role of NN hopping alone in generating topologically distinct bands. In panel (e), adding moderate NNN hopping on top of strong NN hopping opens several mid-band gaps. Interestingly, these are fewer and narrower than in panel (b), where $t_1$ was weaker. This non‐monotonic behavior highlights a subtle competition: large NN hopping can broaden the bands enough to partially suppress the gap-opening influence of NNN hopping. Because band dispersion from large $t_1$ can override some of the localization tendencies of the $t_2$ loops. In panel (f), both hopping channels are at full strength, and the band structure becomes highly deformed. A dense pattern of wide, asymmetric gaps appears across the energy range, and all spectral symmetries are broken. This regime hosts isolated bands, as shown below, with large individual Chern numbers. This is especially promising for realizing topologically robust Hall plateaus with high quantized conductance.

Figure \ref{baand} displays the periodic-boundary‐condition band structures for the 1/5-depleted lattice at six combinations of ($t_1$, $t_2$) when the small plaquettes (area $S$=1) carry the full flux $f$=1. In panel (a), the butterfly is very compressed vertically. Most gaps at high-symmetry points are closed, indicating that with only moderate NN hopping, the small-square geometry (high flux density) does not open wide gaps except near band edges. This reflects limited constructive interference from a single hopping scale. As shown in panel (b), even for modest diagonal hopping, multiple mid-spectrum gaps open that were closed in (a). This shows that adding just one half-strength NNN hopping is enough to significantly alter the band connectivity and to separate formerly overlapping bands. panel (c) shows weak NN and strong NNN hopping opening up gaps even more effectively. Because, the diagonal paths dominate the interference network in the small-cell geometry, producing pronounced gaps. Consequently, the spectrum fragments into well-separated minibands.

Increasing $t_1$ with $t_2=0$ (see panel (d)), compared to (a), widens the overall bandwidth, but, similarly, gaps at many symmetry points remain closed. Only near the top/bottom of the spectrum do small gaps appear. The enhanced NN hopping increases dispersion but still lacks the extra paths needed to stabilize large gaps in this dense flux environment. In panel (e) Compared to (d), mid-spectrum gaps now appear, but fewer than in (b). This illustrates a non-trivial interplay: when $t_1$ is large, adding moderate 
$t_2$ can either open or close specific gaps. This non-monotonic behavior reflects the competition between edge and diagonal hopping phases. In panel (f) the band structure is markedly deformed: several new gaps open throughout the spectrum. The strong NNN hopping introduces additional interference phases, and as will be shown, breaks particle-hole and flux–inversion symmetries and creating new regions of band isolation.

A striking contrast emerges when comparing Figs. \ref{band}(a,d) and Figs. \ref{baand}(a,d), both corresponding to the pure NN-hopping regime ($t_2=0$), but differing in the geometry of flux threading. In the latter case, where flux is threaded through the small plaquettes, the energy bands exhibit extended degeneracy across much of the Brillouin zone, with sharp bifurcations occurring specifically along the high-symmetry line from 
$\Gamma=(0,0)$ to $M=(\pi,\pi)$. This effect is absent in the former case, where flux is threaded through the large plaquettes. The distinction arises from the difference in phase accumulation around magnetic loops: flux through small plaquettes induces stronger local Peierls phases that modulate interference on the scale of nearest-neighbor loops. In this setting, the system retains sublattice symmetry and supports symmetry-protected degeneracies that persist across momentum space due to destructive interference, until they are lifted by constructive interference at specific k-points. In contrast, threading flux through the large plaquettes causes the Peierls phases to spread over longer paths, and no sharp phase cancellation or enhancement occurs along specific momentum-space directions. This breaks momentum-space degeneracies more broadly. As a result, the bands become more dispersive and are typically non-degenerate across most of the Brillouin zone.

It is interesting to note that unlike the spectra shown in Fig. \ref{baand}, which exhibit exact particle–hole symmetry, the band structures in Fig. \ref{band} lack such symmetry. This difference arises from the geometry of magnetic flux threading. In Fig. \ref{band}, as already discussed above, flux penetrates the large plaquettes (area $S=4$), introducing Peierls phases along extended hopping loops that span multiple sublattices. Although the lattice remains bipartite in the absence of diagonal hopping, the phase pattern breaks sublattice (chiral) symmetry in the effective tight-binding Hamiltonian. As a result, the spectrum does not remain symmetric under $E_k\rightarrow -E_k$. In contrast, when flux is confined to the smaller plaquettes, as in Fig. \ref{baand}, the Peierls phases respect the local bipartite structure of the nearest-neighbor network, and particle–hole symmetry is preserved.

\begin{figure*}[th!]
    \centering
   \includegraphics[width=17.5cm]{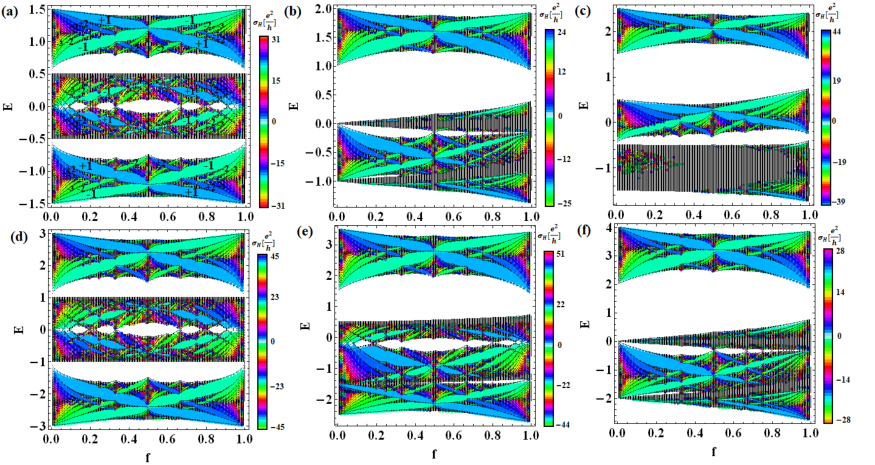}
   \caption{(Color online) Phase diagrams for large squares subjected by magnetic flux at (a) $t_{1}=0.5$, $t_{2}=0$ (b) $t_{1}=0.5$, $t_{2}=0.5$ (c) $t_{1}=0.5$, $t_{2}=1$ (d) $t_{1}=1$, $t_{2}=0$ (e) $t_{1}=1$, $t_{2}=0.5$ (f) $t_{1}=1$, $t_{2}=1$. The color of each gap encodes the value of its associated Chern number. The Chern number $C=0$ is plotted in black and white in the figures.}
   \label{fig2}
\end{figure*}

\subsection{Hofstadter Spectra}

We now present a detailed topological analysis of the Hofstadter spectra for both the large- and small-square configurations, shown in Fig.~\ref{fig1}(a) and Fig.~\ref{fig1}(b), under varying the NN hopping \(t_{1}\) and the NNN hopping \(t_{2}\) parameters. In what follows, we compute the Chern numbers for each energy gap and analyze the evolution of quantized Hall conductivities as functions of magnetic flux and hopping strength.

Figure \ref{fig2} presents Hofstadter spectra for the large-square configuration ($S=4$), showing the energy spectra as functions of magnetic flux $f$, with gaps colored according to their Chern numbers. The top row corresponds to $t_1=0.5$, while the bottom row displays results for $t_1=1.0$, with $t_2=0$, $0.5$, and $1.0$ increasing from left to right in each row.

In panel (a), representing the weak NN regime without NNN hopping, the energy spectrum is narrow and retains exact particle–hole and flux-inversion symmetries. The Hofstadter butterfly is barely resolved, and while only a few small gaps are visible near the band edges, the Chern numbers follow a regular and symmetric sequence. The Chern numbers are paired with opposite signs, so that the total Chern number over all bands vanishes, resulting in zero net Hall conductivity in the bipartite lattice. This configuration serves as a reference point for understanding the effects of symmetry breaking. Introducing moderate NNN hopping in panel (b) immediately distorts the spectrum. Symmetries are broken, and the butterfly becomes more fragmented. Several new gaps open, particularly in the lower part of the spectrum, and an imbalance emerges in the Chern number distribution. A nonzero total Chern number arises, indicating an unbalanced distribution of topological charge. These changes reveal the topological influence of diagonal interference in reshaping the energy landscape, even when NN hopping remains weak. In panel (c), where NNN hopping dominates, the deformation of the butterfly becomes even more pronounced. Mid-spectrum bands shift apart, the central gap becomes well-defined, and small gaps proliferate throughout the energy range. The Chern number distribution becomes increasingly scattered, especially at lower energies, and the total Chern sum over the visible bands remains clearly nonzero. This regime is characterized by frequent topological transitions and the emergence of large Chern numbers driven by strong diagonal coupling. This leads to unconventional topological phases with nontrivial quantized Hall responses.

\begin{figure*}[th!]
    \centering
    \includegraphics[width=17.5cm]{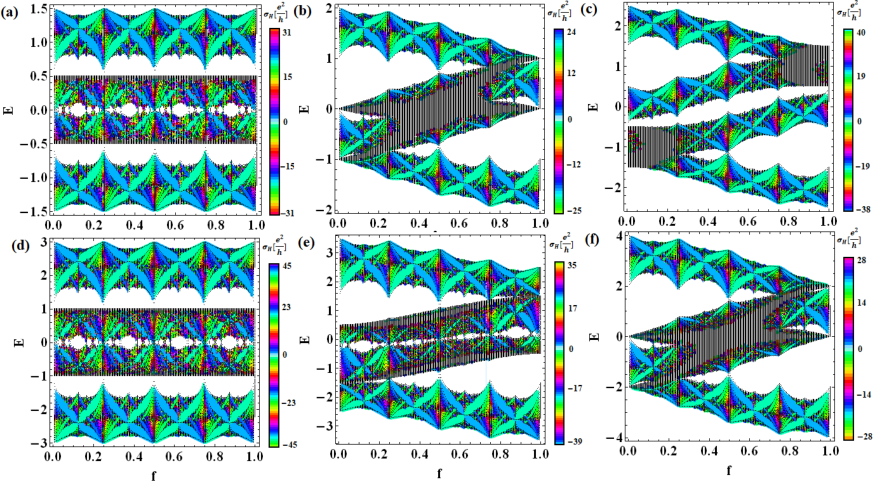}
    \caption{(Color online) Phase diagrams for small squares threaded by magnetic flux for various values of \(t_{1}\) and \(t_{2}\): (a) \(t_{1}=0.5\), \(t_{2}=0\); (b) \(t_{1}=0.5\), \(t_{2}=0.5\); (c) \(t_{1}=0.5\), \(t_{2}=1\); (d) \(t_{1}=1\), \(t_{2}=0\); (e) \(t_{1}=1\), \(t_{2}=0.5\); (f) \(t_{1}=1\), \(t_{2}=1\). The color of each gap encodes the value of its associated Chern number. The Chern number \(C=0\) is plotted in black and white.}
    \label{fig6}
\end{figure*}

Turning to the bottom row, panel (d) explores the case of strong NN hopping without diagonal contributions. The spectrum now spans a wider energy range and reveals a more structured butterfly, with additional gaps appearing near the center. However, since the system still lacks NNN hopping, the particle–hole and flux-reversal symmetries remain intact. In contrast to panel (a) the Chern numbers become larger in magnitude. But they are symmetrically arranged, ensuring that the net Hall conductivity remains zero over the whole spectrum, similar to panel (a). In panel (e), moderate NNN hopping is added to the strong NN regime. The spectrum becomes more irregular, with a broader central gap and many additional smaller gaps emerging at asymmetric flux values. Unlike in panel (d), the Chern numbers no longer cancel pairwise, and finite net Hall conductivity emerges. The interplay between edge and diagonal hoppings introduces new interference paths that break inversion symmetry and produce a stepwise evolution of Chern indices across the spectrum. Finally, panel (f) shows the case where both $t_1$ and $t_2$ are strong. The butterfly is heavily deformed, with clear asymmetries in both energy and flux, and a complex arrangement of gaps across the entire bandwidth. Surprisingly, despite these broken symmetries, the total Chern number remains zero in this case, pointing to a delicate balance between positive and negative topological contributions. This highlights a subtle and nontrivial form of topological compensation, where symmetry is lost but global topological neutrality is preserved.

\begin{figure*}[th!]
    \centering
    \includegraphics[width=17.25cm]{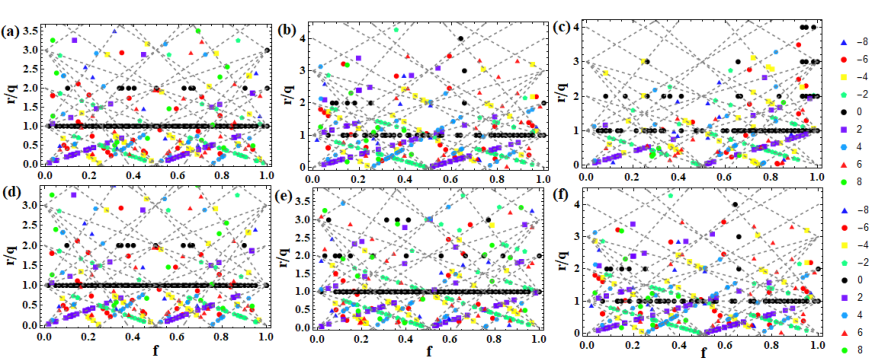}
    \caption{(Color online) Normalized gap index $r/q$ versus flux $f=p/q$ (Landau fan diagrams) for large squares threaded by magnetic flux for different combinations of \(t_{1}\) and \(t_{2}\): (a) \(t_{1}=0.5\), \(t_{2}=0\); (b) \(t_{1}=0.5\), \(t_{2}=0.5\); (c) \(t_{1}=0.5\), \(t_{2}=1\); (d) \(t_{1}=1\), \(t_{2}=0\); (e) \(t_{1}=1\), \(t_{2}=0.5\); (f) \(t_{1}=1\), \(t_{2}=1\). Each colored symbol represents a gap, labeled by the associated Chern number. Straight‐line trajectories reveal the underlying Diophantine relation.}
    \label{fig4}
\end{figure*}

Figure \ref{fig6} displays the Hofstadter spectra for the small-square geometry ($S=1$) under periodic boundary conditions, with each panel showing the energy bands as a function of magnetic flux $f$. The gaps are color-coded by their associated Chern numbers, allowing direct visualization of the topological character of each configuration. The panels are arranged row-wise: the top row corresponds to $t_1=0.5$, and the bottom row to $t_1=1.0$, with $t_2=0, 0.5$, and $1.0$ increasing from left to right across each row.

In the top row (panels (a)–(c)), where NN hopping is weak, the spectrum is initially narrow and exhibits near-perfect symmetry in panel (a) ($t_2=0$). Most gaps are closed except near the band edges, and the Chern numbers appear in symmetric pairs about $E=0$, reflecting the preserved particle–hole and flux-inversion symmetries. As NNN hopping is introduced in panel (b) ($t_2=0.5$), several mid-band gaps open, and the spectrum becomes visibly asymmetric. The Chern numbers no longer cancel perfectly, indicating a nonzero net Hall conductivity across the bands. In panel (c) ($t_2=1.0$), diagonal hopping dominates the dynamics, significantly deforming the butterfly structure. Wide gaps emerge throughout the spectrum, particularly in the central region, and the Chern numbers grow in magnitude and asymmetry. This demonstrates that even in the weak-NN regime, strong diagonal hopping is sufficient to induce topologically nontrivial phases with high Chern numbers.

In the bottom row (panels (d)–(f)), where $t_1=1.0$, the increased NN hopping leads to broader spectra and stronger band dispersions. In panel (d) ($t_2=0$), the spectrum is wider and retains full symmetry, with a regular arrangement of gaps and symmetric Chern number distribution. Adding moderate diagonal hopping in panel (e) ($t_2 =0.5$) disrupts this balance: new gaps appear, some close, and the symmetry about $E=0$ is partially broken. The partial restoration of symmetry compared to panel (d) reflects the competition between dispersive NN terms and topological interference from NNN terms. The Hall conductivity becomes asymmetric. In panel (f) ($t_2=1.0$), both hoppings are strong, and the spectrum is fully deformed. Many gaps carry large Chern numbers, and the lack of symmetry reveals the strong interplay between NN and NNN hoppings. The Hall conductivity becomes symmetric again, exhibiting a non-monotonical behavior. The combined effect of strong edge and diagonal paths yields a highly tunable topological phase diagram, where topological transitions and quantized Hall plateaus can be engineered by adjusting the hopping parameters.

\begin{figure*}[th!]
    \centering
    \includegraphics[width=17.25cm]{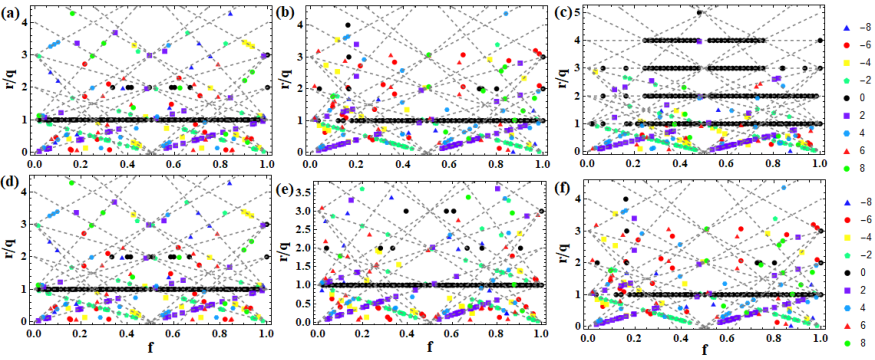}
    \caption{(Color online) Normalized gap index $r/q$ versus flux $f=p/q$ (Landau fan diagrams) for small squares threaded by magnetic flux for different combinations of \(t_{1}\) and \(t_{2}\): (a) \(t_{1}=0.5\), \(t_{2}=0\); (b) \(t_{1}=0.5\), \(t_{2}=0.5\); (c) \(t_{1}=0.5\), \(t_{2}=1\); (d) \(t_{1}=1\), \(t_{2}=0\); (e) \(t_{1}=1\), \(t_{2}=0.5\); (f) \(t_{1}=1\), \(t_{2}=1\). Each colored symbol represents a gap, labeled by the associated Chern number. Straight‐line trajectories reveal the underlying Diophantine relation.}
    \label{fig8}
\end{figure*}

\subsection{Landau Fan Diagrams}

For each energy gap, we solve Eq.~\eqref{eq9} and construct corresponding Landau fan diagrams. We plotted the Landau fan diagrams using the parameters of the Diophantine equation in Fig. \ref{fig4} and Fig. \ref{fig8} for the two models. Each diagram displays the normalized gap index $r/q$ (proportional to the number of filled states) versus magnetic flux $f=p/q$, with straight-line trajectories whose slopes and intercepts encode the Diophantine invariants $C_r$ and $s_r$, respectively. For convenience, we restrict \(C_{r}\) to vary between \(-8\) and \(8\). 

Figure \ref{fig4} presents the Landau fan diagrams corresponding to the large-square geometry ($S=4$) for six representative hopping configurations. In panels (a) and (d), where only NN hopping is present ($t_2=0$), the system exhibits exact particle–hole and flux-inversion symmetries. The fan lines are straight, symmetric about $r/q=1/2$, and correspond to integer Chern numbers that come in canceling pairs, yielding a net zero Hall conductivity over the full spectrum. Increasing the NN hopping from $t_1=0.5$ to $t_1 =1.0$ (comparing panels (a) and (d)) broadens the energy range but preserves the symmetric Diophantine structure.

The introduction of NNN hopping in panels (b), (c), (e), and (f) leads to symmetry breaking and visible deformation of the fan patterns. In panel (b), where $t_2=0.5$, small deviations from perfect line alignment begin to appear—indicating that some gaps are destabilized or shifted due to interference from diagonal hopping paths. In panel (c), where both $t_1$ and $t_2$ are moderate-to-strong, the fans are significantly distorted: lines appear irregular, and clusters of points with zero Chern number proliferate. This reflects complex gap behavior and signals the onset of topological transitions, with a nonzero net Chern number emerging in certain flux intervals.

Panels (e) and (f) illustrate more dramatic effects: fan lines corresponding to large Chern numbers ($C_r=\pm 4,\pm 5$) become prominent, while point scatter increases markedly. The intercepts $s_r$ vary widely, reflecting a shift in the Diophantine character of the spectrum. These patterns confirm that the interplay between NN and NNN hopping generates a rich variety of topological phases. In particular, the emergence of asymmetric fan structures and nonzero total Chern sums demonstrates that the depleted lattice, when perturbed by diagonal hopping, supports robust and tunable Chern insulator states beyond the standard Hofstadter regime.

Figure \ref{fig8} shows the Landau fan diagrams for the small-square geometry ($S=1$) for six representative hopping configurations. In panel (a), where both hoppings are weak and $t_2=0$, the fan pattern is symmetric and sparse. Only a few low-index Chern lines ($C_r =±1,±2$) are visible, and the overall structure reflects the preserved particle–hole and flux-inversion symmetries of the NN-only, bipartite system. The net Chern number vanishes, and all bands appear topologically balanced.

As diagonal hopping is introduced in panel (b) ($t_2=0.5$), the fan begins to deform: certain gaps shift, and lines with higher Chern numbers appear. The asymmetry in the fan pattern signals a breaking of inversion symmetry and the onset of nontrivial topological transitions. In panel (c), where NNN hopping dominates ($t_2=1.0$), the fan becomes much denser and more irregular. High-index Chern lines (e.g., $|C_r|=4,5$) proliferate, and many gaps close or re-open as flux varies, reflected in point scatter and the presence of $C_r=0$ trajectories. These features indicate a fragmented band structure and nontrivial topology even in the weak NN limit.

The second row (panels (d)–(f)) illustrates the effect of increasing $t_1$. In panel (d), with strong NN hopping and no diagonals, the fan structure is again symmetric and sharply defined, with evenly spaced lines of low Chern number. This confirms that strong edge-path hopping alone stabilizes simple topological gaps without inducing net Hall conductivity. Panel (e), with moderate $t_2=0.5$, exhibits both ordered and disordered features: clear low-index lines coexist with scattered higher-index fans. The breaking of symmetry is more pronounced, and the net Chern number can become nonzero across subsets of the spectrum. Finally, in panel (f), where both $t_1$ and $t_2$ are strong, the fan diagram is the most intricate. A multitude of fan lines with a wide range of slopes and intercepts appears, and point scatter is significant. This reflects a highly fragmented Hofstadter spectrum and confirms that the combination of NN and NNN hopping in the small-plaquette regime leads to a dense and tunable landscape of topological phases.

%%%%%%%%%%%%%%%%%%%%%%%%%%%%%%%%%%%%%%%%%%%%%%%%%%%%%%%%%%%%%%%%%%%%%%%%%%%
\section {Conclusion} \label{s4}

We have studied the Hofstadter spectrum and quantum Hall effects in a two-dimensional 1/5-depleted square lattice under a perpendicular magnetic field, considering both NN and NNN hopping amplitudes. Two configurations were analyzed: one where the magnetic flux threads the large square plaquettes of the lattice, and another where it threads the small ones. For both cases, we computed the energy spectra under periodic boundary conditions and identified the evolution of energy bands and band gaps as a function of $t_{1}$ and $t_{2}$. The characteristic Hofstadter butterfly structure was observed, with its shape, symmetry, and gap widths significantly affected by the choice of hopping parameters.

A key result is the strong impact of NNN hopping on both the geometry of the spectrum and the topological character of the system. While the NN-only model preserves particle-hole and flux-inversion symmetry, the inclusion of $t_{2}$ breaks these symmetries, distorts the butterfly, and opens new gaps. Remarkably, we find that the sum of the Chern numbers over the low-energy Hofstadter bands can become non-zero when $t_{2}\neq 1$, revealing topological phases with a net quantized Hall conductance. This nontrivial Chern-number accumulation is highly sensitive to the hopping parameters and highlights the tunability of the model.

We also constructed Landau fan diagrams and analyzed the Diophantine equation $r = p\,C_{r} + q\,s_{r}$, extracting the topological invariants $C_{r}$ and $s_{r}$ for each energy gap. The resulting fan structures exhibit varying degrees of scatter, line density, and symmetry depending on the hopping configuration, offering insight into the stability and structure of topological gaps. 

Overall, our results demonstrate that introducing depletion and diagonal hopping into square lattice models leads to a rich landscape of fractal energy spectra and tunable topological phases. These findings provide a foundation for realizing engineered Chern insulators in cold-atom systems, moiré lattices, and oxide heterostructures based on depleted geometries.

%%%%%%%%%%%%%%%%%%%%%%%%%%%%%%%%%%%%%%%%%%%%%%%%%%%%%%%%%%%%%%%%%%%%%%%%%%%
\section*{acknowledgment}
S.A. and M.V.H. acknowledge the support from the research council of University of Zanjan and Iran Nanotechnology Innovation Council. G.G. gratefully acknowledges funding from the U.S. National Aeronautics and Space Administration (NASA) via the NASA-Hunter College Center for Advanced Energy Storage for Space  under cooperative agreement 80NSSC24M0177.

%%%%%%%%%%%%%%%%%%%%%%%%%%%%%%%%%%%%%%%%%%%%%%%%%%%%%%%%%%%%%%%%%%%%%%%%%%%

\end{document}